\documentclass[a4paper]{jpconf}
\usepackage{graphicx}
\begin{document}
\title{Confusion background from compact binaries}

\author{T.\ Regimbau}
\address{UMR ARTEMIS, CNRS, University of Nice Sophia-Antipolis, \\
Observatoire de la C\^ote d'Azur, BP 4229, 06304, Nice Cedex 4}
\ead{regimbau@oca.eu}

\author{Scott A.\ Hughes}
\address{Department of Physics and MIT Kavli Institute,\\
 77 Massachusetts Avenue, Cambridge, MA 02139}

\begin{abstract}
Double neutron stars are one of the most promizing sources for terrestrial gravitational wave interferometers. For actual interferometers and their planned upgrades, the probability of having a signal present in the data is small, but as the sensitivity improves, the detection rate increases and the waveforms may start to overlap, creating a confusion background, ultimately limiting the capabilities of future detectors.
The third generation Einstein Telescope, with an horizon of  $z > 1$ and very low frequency ``seismic wall'' may be affected by such confusion noise.  At a minimum, careful data analysis will be require to separate signals which will appear confused.  This
result should be borne in mind when designing highly advanced future
instruments.
\end{abstract}

\section{\label{sec:intro}Introduction}

The coalescence of two neutron stars (BNS), two black holes (BBH) or a neutron star and a black hole (NS-BH),
are among the most promising sources for ground-based gravitational wave detectors
due to the huge amount of energy emitted in the last phase of their inspiral trajectory.
The waveform is very well modeled until the last stable orbit
and a detection would provide strong constraints on the source parameters.
With the third generation interferometer Einstein Telescope,
the horizon of compact binaries is expected to reach cosmological distances where it may become  possible to study cosmology. In particular, double neutron stars or neutron star-black holes, if associated with an electromagnetic counterpart, may be used as standard sirens to constrains dark energy \cite{hol05}.
At such distances however, it is likely that the sources create a confusion foreground,
where the detection of a single coalescence becomes difficult and requires advanced data analysis methods \cite{reg09}.
In the first section, we derive the coalescence rate of BNS; in the second section we discuss the different detection regimes and the consequences for Einstein Telescope; in the third section we present our conclusions and the work in progress.

\section{\label{sec:rate}Coalescence Rate}
The final merger of a compact binary occurs after two massive stars in a binary system have collapsed to form neutron stars or black holes and have inspiralled through the emission of gravitational waves. The cosmic coalescence rate is given by:
\begin{equation}
\dot{\rho}_c^o(z) \propto \int \frac{\dot{\rho}_*(z_f)}{1+z_f}P(t_d)dt_d \,\ \mathrm{with}\,\ \dot{\rho}_c^o(0)=\dot{\rho}_0
\end{equation}
In this expression, $\dot{\rho}_c$ is the star formation rate in M$_\odot$Mpc$^{-3}$yr$^{-1}$, $z_f$ the redshift at the time of formation of the binary system and $P(t_d)$ the probability distribution of the delay between the formation and the coalescence.The coalescence rate per interval of redshift $dR^o/dz(z)$ is obtained by multiplying by the element of comoving volume $dV/dz(z)$.
Here, we assume a distribution of the form $P(t_d) \propto 1/t_d$ with $t_d>t_{\min}$ and take $t_{\min}=20$ Myr as representative of BNSs. We use the star formation rates of \cite{hop06,far07,wil08,nag06}, and consider local rates in the range $0.01-10$ Mpc$^{-3}$Myr$^{-1}$,  with reference models of 1 and 0.4, corresponding to the most current estimates derived from statistical studies \cite{kal04} and population synthesis \cite{dfp06,sha08}.

\section{\label{sec:regime}Detection Regime}
The contribution of BNS to the instrumental data falls into three statistically very different regimes, depending on the duty cycle (or the average number of sources present at the detector at the same time):
\begin{equation}
\Lambda(z)=\int_0^z \tau^o(z)\frac{dR^o}{dz}(z) dz
\end{equation}
where $\frac{dR^o}{dz}(z)=\dot{\rho}^o(z) \frac{dV}{dz}$ is the coalescence rate per interval of redshift, and $\tau^o(z)$ is the typical duration of the inspiral in the detector frequency band, which depends strongly on the low frequency limit of the instrument and can last from a few minutes for advanced detectors with $f_L=10$ Hz to a few days for the Einstein Telescope with planned low frequency bound between $1-5$ Hz (see Table~\ref{table-transition} )

\begin{enumerate}

\item {\it Shot noise ($\Lambda<<1$)}: This case describes when the number of sources
is small enough that the interval between events is long compared to
an individual event's duration.  Measured waves are separated by long
stretches of silence and can be resolved individually.  This case
pertains to instruments that are only sensitive to events at low
redshift.

\item {\it Popcorn noise ($\Lambda \sim 1$)}: As the reach of instruments increases, the
time interval between events may come closer to the duration of single
bursts.  Events may sometimes overlap, making it difficult to
distinguish between them.

\item {\it Gaussian ($\Lambda>>1$)}: For instruments with very large reach and
excellent low frequency sensitivity, the interval between events can
be small compared to the duration of an event.  The signals overlap to
create a confusion noise of unresolved sources.
\end{enumerate}

\begin{figure}
\centering
\includegraphics[angle=0,width=0.8\columnwidth]{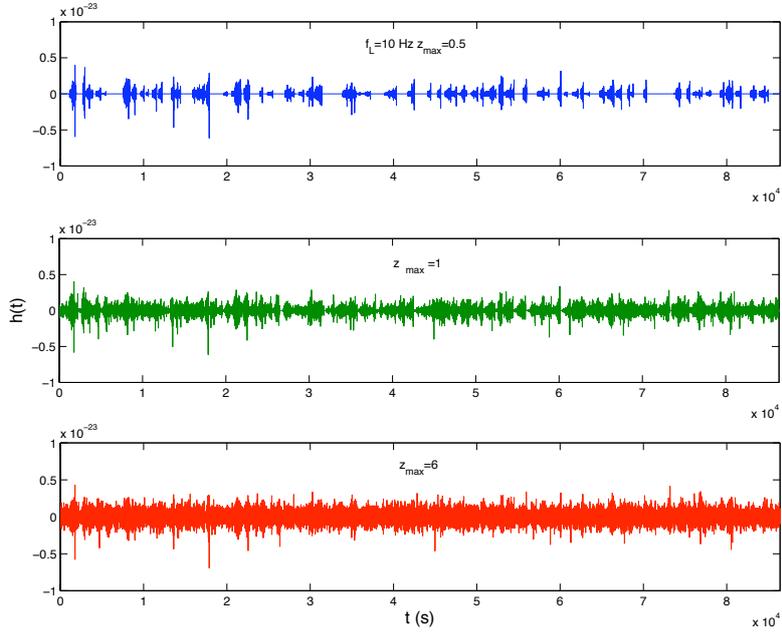}
\caption{Evolution with the detector horizon $z_{\max}$ of a GW time serie  from a simulated population of BNSs. Here we have assumed detector a low frequency bound $f_L=10$ Hz.}
\label{fig-series1}
\end{figure}

\begin{figure}
\centering
\includegraphics[angle=0,width=0.8\columnwidth]{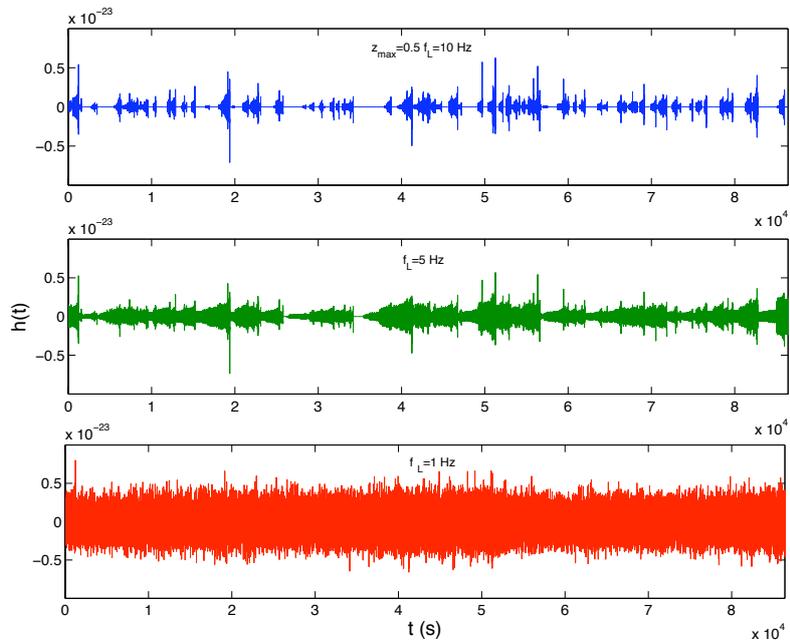}
\caption{Evolution with the detector low frequency bound, of a GW time serie of the from a simulated population of BNSs. Here we have assumed a detector horizon $z_{\max}=0.5$.}
\label{fig-series2}
\end{figure}
\begin{table}
  \centering
  \caption{Threshold between resolved and unresolved BNSs
for different estimates of the source
rate $\dot\rho_c^{\rm o}$ and detector lower frequency bound $f_L$. No
value means that the number of sources at the detector is always $<1$ or $<10$.
$\tau_0$ refers to the typical duration of a source located at $z=0$.}\label{table-transition}
  \begin{tabular}{|l|cccc|}
\hline
$f_L$ & $\tau_0$ & $\dot\rho_c^{\rm o}$ &  $z_*$ & $z_{**}$ \\
\hline
\hline
  10 & 16.7 m & 0.01 & - & -\\
     &        & 0.4 & 0.8-0.9 & -  \\
     &        & 1 & 0.5-0.6 &  $>2$\\
     &        & 10 & 0.2 & 0.5-0.6\\
\hline
   5 & 1.8 d  & 0.01 & - & - \\
     &        & 0.4 & 0.4 & 1-1.2\\
     &        & 1 & 0.25 & 0.6-0.7\\
     &        & 10 & 0.1 & 0.25 \\
\hline
   1 & 5.4 d  & 0.01 & 0.3 & 0.8\\
     &        & 0.4 & 0.08 & 0.2\\
     &        & 1 & 0.06 & 0.13\\
     &        & 10 & 0.03 & 0.06\\
\hline
  \end{tabular}
\end{table}

The limit of the popcorn and the Gaussian regimes, defined by $\Lambda(z_{*})=1$ and $\Lambda(z_{**})=10$ are given in Table~\ref{table-transition} \cite{reg09} for different values of the low frequency bound and local coalescence rate.
With actual and advanced interferometers, whose horizon is a tens or a hundreds of Mpc, we are able to probe only the low duty cycle regime, where sources don't overlap. With the  third generation Einstein Telescope, on the other hand, the horizon is expected to extend to redshift larger than $z>1$, and the signal may fall deep inside the confusion regime, especially between $1-5$ Hz, where it can last for a few days (see Table~\ref{table-transition}).

We find that if the low frequency bound is at $f_L = 1$ Hz, the transition to a Gaussian background at $z_{**}$ is well within the ET horizon; if $f_L = 5$ Hz, it most likely occurs at $z_{**} \sim 0.6 - 1.2$, but can fall
beyond the detection horizon if $\dot\rho_c^{\rm o} <0.15$ Myr$^{-1}$
Mpc$^{-3}$. The transition to a popcorn background at $z_*$, on the other hand, always occur before the detection horizon, and more likely around $z_* \sim 0.25-0.4$, unless for the most pessimistic coalescence rates ($\dot\rho_c^{\rm o}
<0.015$ Myr$^{-1}$ Mpc$^{-3}$).

\section{Conclusions and future work}

With the first and second generations of gravitational-wave detectors the chance of detection is limited by the instrumental noise, but as the sensitivity and the number of sources increase, the waveforms may start to overlap, creating a confusion background, ultimately limiting the capabilities of these instruments.
Experience from
the Mock LISA Data Challenges \cite{bab08} and ideas developed for
the Big Bang Observatory \cite{cut06} prove that disentangling
multiple signals in a gravitational-wave detector's data stream is
certainly possible.
We have developed a Monte Carlo simulation code to generate the signal from populations of extra-galactic binaries in the output of detectors, which can be used to study the statistical properties of the background and investigate advanced data analysis strategies.
The evolution of the signal with the detector horizon and with the low frequency bound is illustrated by the simulated time series in Fig. ~\ref{fig-series1} and Fig. ~\ref{fig-series2}.
One can imagine applying a high pass filter to identify and subtract the sources, iteratively from lower to higher redshifts, from the integrated signal. This work is in progress and will be reported in a future paper.

\end{document}